\documentclass[11pt,a4paper]{article}

\renewcommand{\author}{Emil Khalisi}
\newcommand{\titel}{Eclipses in the Aztec Codices}
\newcommand{\version}{Version 1.63}
\renewcommand{\date}{20th August 2020}

\usepackage{hyperref}
\hypersetup{pdfauthor={Emil Khalisi}}
\hypersetup{pdftitle={\titel }}
\hypersetup{pdfsubject={Online publication, \version of \date }}
\hypersetup{pdfkeywords={Eclipses, Aztec, Mesoamerica, Astronomical dating,
    Tenochtitlan, pre-hispanic, Mexico, Bilimek, Moctezuma, Montezuma, Comet}}

\usepackage[T1]{fontenc}        
\usepackage{mathptmx}           
\usepackage{pifont}             
\usepackage[scaled=0.92]{helvet} 

\usepackage[british]{babel}     
\usepackage{amsmath}            
\usepackage{amssymb}            
\usepackage{microtype}          
\usepackage{paralist}            
\usepackage{arydshln}           
\usepackage{rotating}           

\usepackage{calc}
\usepackage[a4paper]{geometry}  
\usepackage{scrextend}           
\changefontsizes{10pt}
\usepackage{titlesec}
\titleformat*{\section}{\large\bfseries}
\titleformat*{\subsection}{\normalsize\bfseries}

\usepackage{graphicx}           
\usepackage{float}              
\usepackage{caption}            
\usepackage{fancyhdr}
\renewcommand{\headrulewidth}{0.4pt}
\usepackage{colortbl}             
\definecolor{grey20}{RGB}{208,208,208}

\begin{document}


\fancyhead{}
\fancyhead[LO]{%
   \footnotesize \textsc{In original form published in:}\\
   {\footnotesize Habilitation at the University of Heidelberg }
}
\fancyhead[RO]{
   \footnotesize {\tt arXiv: 0000.00000 [physics.hist-ph]}\\
   \footnotesize {Date: 20th August 2020}%
}
\fancyfoot[C]{\thepage}

\renewcommand{\abstractname}{}

\twocolumn[
\begin{@twocolumnfalse}

\section*{\centerline{\LARGE \titel }}

\begin{center}
{\author \\}
\textit{D--69126 Heidelberg, Germany}\\
\textit{e-mail:} \texttt{ekhalisi[at]khalisi[dot]com}\\
%
\end{center}


\vspace{-\baselineskip}
\begin{abstract}
\changefontsizes{10pt}
\noindent
\textbf{Abstract.}
This paper centers on the collection of accounts on solar eclipses
from the era of the Aztecs in Mesoamerica, about 1300 to 1550 AD.
We present a list of all eclipse events complying with the
topological visibility from the capital Tenochtitlan.
Forty records of 23 eclipses entered the various Aztec manuscripts
(codices), usually those of large magnitude.
Each event is discussed with regard to its historical context,
as we try to comprehend the importance the Aztecs gave to the
phenomenon.
It seems that this culture paid noticeably less attention to
eclipses than the civilisations in the ``Old World''.
People did not understand the cause of it and did not care as much
about astronomy as in Babylonia and ancient China.
Furthermore, we discuss the legend on the comet of Moctezuma II.
It turns out that the post-conquest writers misconceived what the
sighting was meant to be.

\vspace{\baselineskip}
\noindent
\textbf{Keywords:}
Solar eclipse,
Aztec,
Mesoamerica,
Astronomical dating,
Moctezuma's Comet.
\end{abstract}

\centerline{\rule{0.8\textwidth}{0.4pt}}
\vspace{2\baselineskip}

\end{@twocolumnfalse}
]




\section{Introduction}

The Mesoamerican peoples developed their culture independent from
the civilisations in the Euro-Asian domain.
Among the tribes on both American continents only the Maya
achieved an advanced level of worldwide recognition.
Of course, each culture bases its merits on forerunners, and so
did the Maya, but they went further than any of the predecessors
and contemporaries.
The period of their flourishing spans from 250 to 900 AD.
Thereafter other peoples soared:
Toltec, Mixtec, Zapotec, Chichimeca, etc.
The last empire before the arrival of the Europeans in 1519 was
established by the Aztecs.
They can be considered as the principal heirs of the Maya, though
their legacy manifests itself significantly less developed.
The destruction of the Aztec kingdom followed in August 1521.

When getting involved into the scientific achievements of the
Aztecs, it comes to light that the knowledge about celestial
happenings remained on a rather low level.
We find no systematic observations on the course of planets, no
star maps, no profound models about the cosmic structure, no
attempts of computing a cycle.
Mathematics did not go beyond a rudimentary stage, and basic
periods were hardly trailed.
The only cycle they cared for was the calendar.
Even this one was not their own invention, but commonly used by
other peoples in Mesoamerica.
It is a matter of viewpoint whether this deficiency of technological
progress is owed to a lack of curiosity or inability.
A nation not producing mathematicians and scientists falls behind
quickly.
The Aztec culture exhibits some parallels to the Romans in the
Mediterranean:
both peoples were very superstitious, but their main concern was
warfare and military power rather than the exploration of the
world they lived in.
Nonetheless, astronomical observations were a constant necessity,
because the belief was interlocked with some repeating phenomena,
for instance, the end of the calendrical cycle, rebirth of the sun,
or re-appearances of the planet Venus.

Most reliable information about natural phenomena dates back to
the last 50 years before the conquest.
Older information is enshrouded in a legendary style, and there
is suspicion that parts of the history were distorted and
re-adjusted over time.
This conjecture is based on the lack of accounts on striking
astronomical phenomena.
Nevertheless, there are also indications that the cultural
heritage was handed down surprisingly correct.


In this paper we deal with astronomy during the Aztec period.
After reviewing some few historical basics, we will focus on solar
eclipses in the manuscripts available.
We present a full list of any event that could have principally
been registered from central Mexico.
Many eclipses turn out mismatched, while others of large magnitude
are absent.
The magnitude (mag) of a solar eclipse is defined as the ratio of
the apparent angular diameter of the obscuring moon,
$\theta_{\rm M}$, to the diameter of the sun, $\theta_{\odot}$,
both of which are approximately 0.5$^{\circ}$ depending on their
exact distance from the observer:
\begin{equation}
\text{mag} = (\theta_{\odot} + \theta_{\rm M} - \Delta ) / 2 \; \theta_{\odot} ,
\end{equation}
with $\Delta$ the distance of the centers of the two disks during
the ongoing eclipse.
Totality is achieved for mag $>$ 1.
The magnitude is not to be confused with ``obscurity'' which is
the fraction of the overlapped area of the two disks
\cite{finsternisbuch}.
An eclipse is not necessarily observed, for the lighting conditions
will start changing, if the magnitude exceeds a value of
$\approx$0.75 or so.
Bad weather and clouds can easily make an eclipse pass without
notice unless it becomes totally dark.
Thus, a high-magnitude obscuration is no guarantee that the event
was actually perceived.

The tracks of historical eclipses are shifted in longitude against
a constant rotation of the earth due to its long-term deceleration.
This displacement is characterised by $\Delta T$, the difference
between a perfectly uniform time and the civil time.
For the era of the Aztecs, $\Delta T$ is well-known from various
astronomical measurements from medieval Europe, Arabia, and China.
Hence, we are on safe ground when presenting eclipse data.
We base our investigation on the \textit{Five Millennium Canon of
Eclipses} by Fred Espenak \cite{espenak}.


\section{Aztec Literature}

Our knowledge about the Mesoamerican peoples is generally poor.
On one hand, plenty of precious documents were destroyed by European
conquerers, on the other hand the natives developed a different way
of communication.
The Aztecs used to sketch a message in the shape of a pictograph.
This way of ``writing'' did not arrive at an abstract usage of
characters or syllables, as the stylised figures were not meant
to reflect the spoken language.
In our time, an issue has to be guessed by discerning fine details
in the glyphs.
This is an important point and should be kept in mind, as the entire
method would be prone to misinterpretation and confusion.
On the whole, we know about the indigenous peoples of pre-columbian
Mexico very much less, though temporally closer, than we do about
the more ancient civilisations of Mesopotamia or Asia.

From the pre-hispanic time only very few manuscripts survived ---
less than two dozens, and they also disappeared over time.
During the first decades following conquest in 1521 historical
annals were recreated by missionaries.
These books are called ``Aztec Codices''.
Table \ref{tab:codices} specifies those codices we trawled through
in search for eclipses.
Some Spanish writers made an attempt to portray the former native
culture as both pictorial copy of the originals and written
descriptions of what the painted images were meant to be.
Thus, almost all preserved manuscripts are post-conquest works.
In some few cases the writers were natives grown up and educated
bi-lingually after the destruction of the Aztec Empire.
None of the documents covers the Aztec history in full, but
they rather focus on events that affected a particular city.
Major episodes such as a drought can be usually found in multiple
sources permitting a rough reconstruction of the political
interconnections.

\begin{table}[t]
\caption{Aztec codices deployed in this paper.}
\label{tab:codices}
\centering
\begin{tabular}{lcl}
\hline
\rowcolor{grey20}
   Codex        &    Year      & Author \\
\hline
   Aubin        &  1576--1608  & (from Tenochtitlan) \\
   Azcatitlan   &  ca.\ 1530?  & indigenous? \\ 
   Borgia       & pre-conquest & Mexican native \\ 
   Chimalpahin  & 1600\dots 1620?& Mexican native \\ 
   Chimalpopoca & post-conquest & unknown \\ 
   Duran        &  ca.\ 1580   & Diego Duran \\
   Florentine   &  1545--1590  & B.\ de Sahagun \\
   Huichapan    &  ca.\ 1632   & J. de San Francisco \\ 
   Mendoza      &  ca.\ 1541   & various \\ 
   Mexicanus    &  ca.\ 1590   & Mexican native? \\ 
   Rios/Vaticanus A&1546--1560? & Pedro de los Rios \\ 
   Telleriano-R.& post-conquest & various \\
   Tlaxcala     &  1581--1584  & Diego M. Camargo \\ 
   Torquemada   &  ca.\ 1615   & J.\ de Torquemada \\
   Tovar/Ramirez&   1579?      & Juan de Tovar \\            
\end{tabular}
\end{table}

The \textit{Codex Borgia} is believed to be one of very few
surviving documents from before the Spanish conquest, but it may
also be a younger copy of a pre-columbian document.
The \textit{Codex Telleriano-Remensis} could be another example
for a copy of an Aztec original.
The name comes from the later owner, the archbishop Le Tellier of
Reims in France, who possessed it in the late 17th century.
We will abbreviate it to ``Telleriano'' henceforth.
The author of this manuscript is unknown,
but it was produced on European paper in the 16th century.
The publisher of the recent facsimile, Eloise Qui\~{n}ones Keber,
has identified at least two different artists and six different
annotators as having worked on it \cite{quinones}.
It harbours natural phenomena like earthquakes, droughts, storms,
and eclipses.

The \textit{Codex Mexicanus} describes, among other codices, the
history beginning with the migration from Aztlan, the ancestral
home of the Aztecs \cite{diel_2018}.
Much has been speculated about the possible location of Aztlan.
Historians tend to place it either to northwestern Mexico near
the Gulf of California or to the southwest US.
We know that once a major society, called Hokoham, existed in the
area of Phoenix in Arizona/USA between 300 and 1500 AD.
That place at Gila River was proposed by the Mexican writer
Francisco Clavijero (1731--1787), then challenged by others
\cite{gallatin_1845}.
We do not intend to draw a link to the migrants who would become
the Aztecs later, because we do not know better.
The name ``Azteca'' is the native word for ``people from Aztlan''
used by several nomadic groups inhabiting the Valley of Mexico
prior to the arrival of the foreigners.
Those nomads called themselves ``Mecitis'' or ``Mexica''
\cite{berdan-anawalt}.
The \textit{Codex Mexicanus} contains calendrical and astrological
information, some of which is related to the practice of medicine.

A very important work is the \textit{Florentine Codex} by the
Spanish friar Bernardino de Sahag\'un (1499?--1590)
\cite{lockhart}.
He journeyed to Mexico in 1529 and spent about 50 years studying
the indigenous culture.
He worked on his books until his death and produced 2,400 pages
facilitating deeper insight to the society before the invasion.
Organised into twelve books, the codex tells about religion,
mythology, and traditional life of the Aztecs.

Today, some manuscripts are scanned and viewable on the web page
of the Foundation for the Advancement of Mesoamerican Studies,
\texttt{www.famsi.org},
or in the World Digital Library, \texttt{www.wdl.org} .
We base the lion's share of our analysis on these codices, and
complement it with other documents wherever necessary.
Unfortunately, Spanish publications are not accessible to the
author, and a lot of valuable information may not be included here.

%
\fancyhead{}
\fancyhead[CE, CO]{\footnotesize \itshape E.\ Khalisi (2020): \titel}
\renewcommand{\headrulewidth}{0pt}


\section{Perception of Time}

The original Aztec chronicles were constructed such that they
featured one or two events as being significant for one special
year.
This way of recording someone's own history resembles the Indian
tribes of North America who administered their past in so-called
``wintercounts'':
a spiral of icons was penned on buffalo leather to memorise the
incident.
A person in charge learnt the icons by heart and was able to
recount a story to any image, e.g.\ battles, floods, death of
rulers, etc.

The history of the Aztec people follows a timeline subdivided
into cycles of 52 years.
Each year within the cycle is built from names and numericals in a
double arrangement, e.g.\ ``8 Reed''.
It originates from a day-number combination of two calender types,
a 260-day- and 365-day-calendar.
The name of the solar year is identical with the last ``regular''
day at the end of that particular year (360th day excluding the
five extra days to complete it).
For the basics of the calendar we refer to the elementary textbooks,
e.g.\ \cite{kelley-milone}.
The full round of 52 years is what the Aztecs called a ``century''.

Every century was marked by the ``New Fire Ceremony'', a religious
festival to ensure the movement of the cosmos and the rebirth of
the sun.
The ceremony was not held immediately at the beginning of the
52 years but at the end of its first year,
i.e.\ between the year designations ``1 Rabbit'' and ``2 Reed''.
The delay may be the result of a calendar reform in 1506, because
in earlier times the first year, 1 Rabbit, was marked by droughts
and famine.
This view about the 1-Rabbit-years was disseminated by Sahagun
telling that the Mexicas feared them of bad fortune.
The ancestors from time immemorial would have declared such years
dangerous because of floods, eclipses, and earthquakes.
The global destruction would come up as an option at the completion
of each century.
Thus, the Aztecs performed great sacrifices to their gods, and,
when the precise day arrived, did penance and abstained from
misbehaviour.
Then they extinguished all lights and fires until the day ended
and lit new fires \cite{berdan-anawalt}.
A night vigil was kept watching the stars to pass certain marks.
If they did, a new ``contract'' with the gods was signed;
if not, the sky would stop turning around and the sun would not
re-appear.
People got prepared for the end of the world each time when the
century ceased.

The festivals are most likely a matter of fact, however, the
Mexica's fear from the so-called ``1-Rabbit''-years is not
asserted by all Spanish writers.
During the early years of contact there was a lot of
misunderstanding and probably deliberate transformation of the
native practices into a desired world view of the occupants.
It is the typical course of history that all representatives of
any religion aimed at abolishing of ``pagan'' rites in the
conquered land to be replaced by their belief.
Therefore, various old customs were deliberately cast into a
questionable light by later writers.

Many important events in the history of the Aztecs are geared to
the year 1 Flint (others call this sign ``Knife'').
It marks the middle of the 52-year cycle.
The name constitutes a historical peg to emphasise the ``cosmic
validity'' of relevant political stages.  
In accord to the style of cyclic history, time itself is said to
have started in a year bearing that name.
The migration from Aztlan began in 1168
(others propose 1197 \cite{milbrath_1997});
it ended after exactly 3 calendric cycles in 1324;
the enthronement of the first king took place in 1376;
independence was gained in 1428;
and so on.
All these years correspond to 1 Flint (except 1197).
As in any other culture, there is a sense of sacred balance to the
story, so, not every date will be conform to reality.

Legend has it that the capital Tenochtitlan was founded in 1325
after more than 150 years of wandering around
(see paragraph on page \pageref{ch:tenochtitlan}).
A chief called Tenoch (1299--1375) is said to have been elected
to power by the council of elders and ruled for about 51 years
or so \cite{gallatin_1845}.
Tenoch's successor is regarded as the first official ruler who
renamed the city in honour of the former chief.
The military power of the Aztecs commenced when a triple alliance
was formed with two other cities, Texcoco and Tlacopan.
Then the might spread throughout much of the central and southern
Mexico, drawing sustained tribute from the vanquished vassals.
Because of this, the Aztecs were hated among the tribes gladly
helping the Europeans to shake off the yoke.
What followed was a complete destruction of all native cultures
and an oppression even worse by the new masters.


\section{Mythological Concept of Eclipses}

\begin{figure*}[t]
\includegraphics[width=\linewidth]{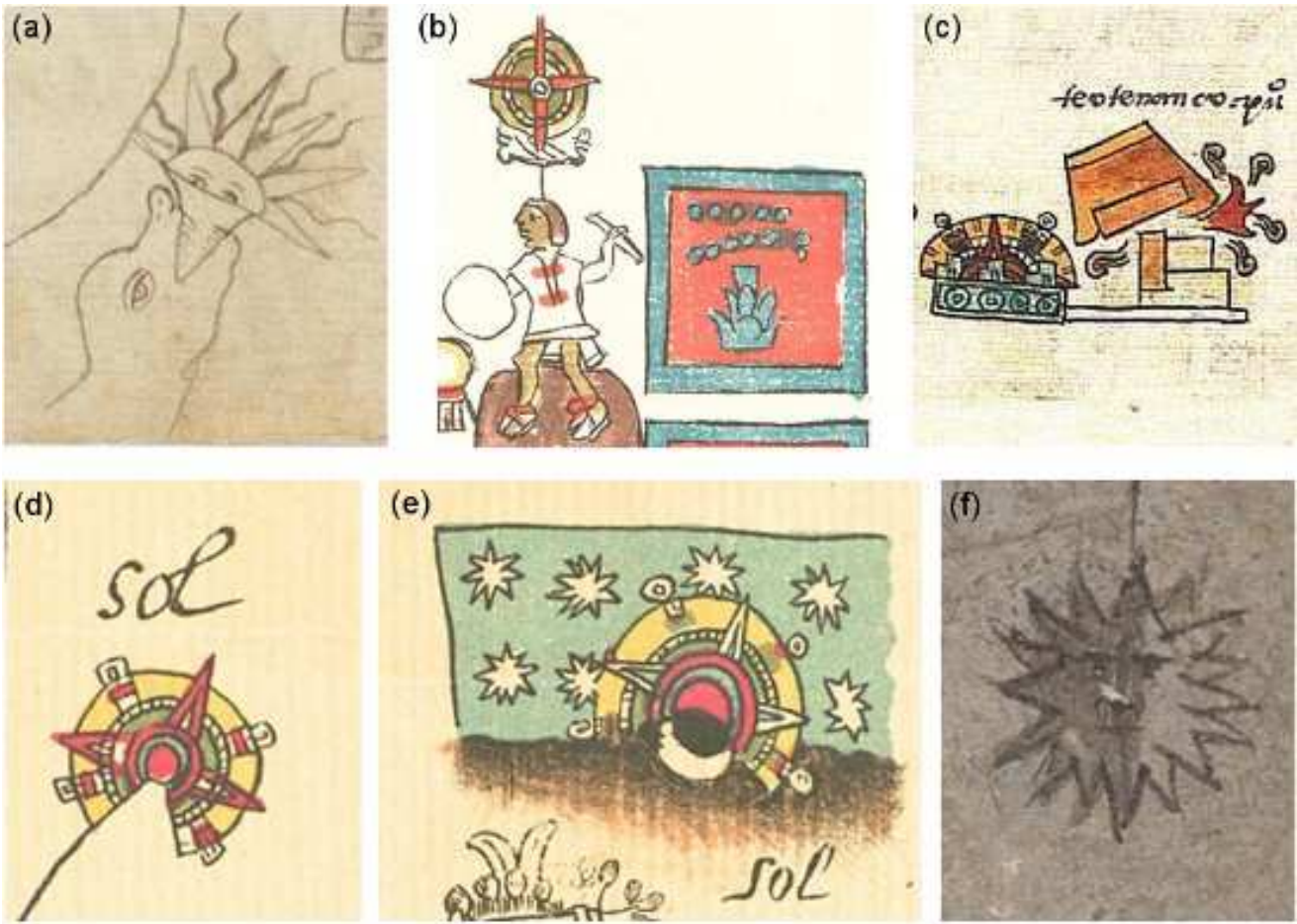}
\caption{Various glyphs for eclipses from the codices.
     (a) \textit{Azcatitlan}, 7v: eclipse of 1301, 1303, or 1311.
     (b) \textit{Vaticanus A}, 76r: eclipse of 1437.
     (c) \textit{Mendoza}, 10r: eclipse of 1477.
     (d) \textit{Telleriano}, 37r: eclipse of 1477.
     (e) \textit{Telleriano}, 40v: eclipse of 1496.
     (f) \textit{Mexicanus}, p56: eclipse of 1524(?).}
\label{fig:eclipseglyphs}
\end{figure*}

The glyphs for solar eclipses vary in different codices
(Fig.\ \ref{fig:eclipseglyphs}).
The sun was usually depicted as a circle of concentric rings with
outstretched spikes for its rays.
As for an eclipse, it was truncated and dot-like circles (stars)
attached.
Since stars will not be visible in a partial eclipse, this glyph
does not tell much about the particular event.
The Aztec sign just applied to any kind of eclipses of the sun.
For us, it seems impossible to retrieve more detailed information
neither on totality, nor size of obscuration,
nor the time of day, nor the season of the year.
Because eclipses and other astronomical events are amply available
from Europe, Arabia, or China, they do serve for a reconstruction
of the circumstances as well as geophysical effects.
In contrast, the Mesoamerican information precludes usefulness for
scientific analysis, e.g.\ to precisely adjust the geographical
position of the observer.

As much as we know today about the peoples in old Mexico, they
had a variety of mythological concepts about eclipses.
We present three examples.
The first and easiest is given by the anthropologist Eduard Seler
(1849--1922) referring to the \textit{Codex Vaticanus}
\cite{seler_1902}:
the Aztecs thought that a jaguar was going to eat the sun.
The jaguar was the pictoglyph of the 14th day in the calendar as
well as the symbol of darkness.
Of course, every eclipse was awe-inspiring, and people used to make
as much noise as possible to frighten away the monster to let the
sun go.


A completely different idea was that an ``underworld'' was located
in the sky \cite{kelley-milone}.
Because of the shining sun, it could not be seen.
Only at eclipses a glance would be granted into the land of the
dead.
Planets and stars can go there from time to time, and they challenge
the sun.
For example, Mercury and Venus were considered twins.
Their common feature was the ability of approaching the hot
luminary extremely close.
At times of conjunction they demonstrated their prowess to the
lords of the dead:
they performed ``a dance in the flames''.
A couple of days later they would reappear in the sky unharmed.
This dance is seen for a moment during eclipses.

A third belief concerned demonic creatures of darkness.
They are depicted with banded faces wielding weapons, and they
have their hair pulled into two hornlike projections.
These demons, named ``Tzitzimime'', were sky-dwelling skeletal
beings corresponding to the souls of sacrificed warriors.
Information about them is handed down in the
\textit{Florentine Codex} by Sahagun in Books 7 and 8.
Karl Taube, who studied these figures in more detail, gave a
number of examples for their appearance \cite{taube_1993}.
Two of the warriors are positioned to the left and right of the
sun in Figure \ref{fig:bilimek}.
However, the identification of special depictions is debated
among scholars: 
some historians put them on a level with goddesses of war and
suggest various names,
others with supernatural beings,
or they may stand for planets or star constellations.
One study links them with the black disk of the New Moon
becoming visible at a solar eclipse \cite{milbrath_1995}.
Whatever the Tzitzimime are to represent, they are said to descend
down to Earth during totality
as well as other periods of darkness and destroy the world.
This would bring about the end to mankind.
Therefore, an extinguishing sun was always feared as the end of
the world.

It is noteworthy that the Aztecs did not relate an eclipse with
the death of an individual king or warfare.
This contradicts the statement by Susan Milbrath
\cite{milbrath_1995} who tried to make similar connections as they
are commonly known from the ``old world''.
Among natural disasters the eternal darkness and the appearance of
these Tzitzimime was taken serious but not understood as an
announcement of upcoming floods or so.
In spite of their superstitious attitude the Aztecs distinguished
between events in the sky and events on earth.


\section{Records of Solar Eclipses}

\begin{table*}[t!]
\caption{Data for all solar eclipses over Tenochtitlan
    (19$^{\circ}15^{\prime}$ N, 99$^{\circ}6^{\prime}$ W) between 1300
    and 1550 AD.
    Local time (LT), magnitude, and altitude of the sun refer to
    the instant of maximum eclipse phase.
    The star (*) in the type-column denotes the path passing centrally
    over the city. --- 
    Additionaly, in the upper part of the Table: some eclipses in ``Aztlan''.}
\label{tab:eclipses}
%
\changefontsizes{9.5pt}
\centering
\begin{tabular}{l|r|c|r|c|r|p{3.8cm}|p{2.9cm}}
\hline
\rowcolor{grey20}
Date         & Aztec year & Type &\multicolumn{1}{c|}{\cellcolor{grey20}LT}
                                    & Magn. & Alt. & Codex     & Ref. \\
\hline
1155, Jun 01 &  1 Reed  & A & 16:47 & 0.800 & 26.5 & (Aztlan)  & Fig.\ \ref{fig:aztlan} \\
1156, Nov 14 &  2 Flint & A & 16:19 & 0.945 &  6.0 & (Aztlan)  & Fig.\ \ref{fig:aztlan} \\
1165, Nov 05 & 11 House & A & 10:46 & 0.797 & 36.7 & (Aztlan)  & Fig.\ \ref{fig:aztlan} \\
\cdashline{1-8}[0.5pt/5pt]     
1196, Sep 23 &  3 Flint & T & 12:35 & 0.965 & 52.0 &             & Fig.\ \ref{fig:aztlan} \\
1198, Feb 07 &  5 Rabbit& A & 17:43 & 0.269 &$-$0.3& Mexicanus, p29 &
                       $\;\;$\begin{rotate}{90}$\Lsh$\end{rotate} mistaken?
                         \cite{aveni_1999} \\ 
1199, Jul 24 &  6 Reed  & A & 15:03 & 0.485 & 47.5 &             &
                               Fig.\ \ref{fig:comet1199}; E.K. \\
1200, Jul 12 &  7 Flint & A & 16:16 & 0.166 & 33.5 &             &    \\
1203, May 12 & 10 Reed  & T & 14:26 & 0.353 & 54.0 &             &    \\
1205, Sep 14 & 12 House & T & 12:01 & 0.504 & 57.2 &             &    \\
%
\hline
%
1301, Feb 09 &  4 House & A*&  9:49 & 0.975 & 42.3 &  
                       Azcatitlan, Plate 9 &
                       Fig.\ \ref{fig:eclipseglyphs}a; \cite{aveni_1999} \\
1303, Jun 15 &  6 Reed  & H & 18:47 & 0.980 &$-$2.3&           &
                       (Fig.\ \ref{fig:eclipseglyphs}a?) \\
1307, Sep 27 & 10 Reed  & A &  6:57 & 0.478 & 14.5 &           &      \\
1308, Sep 15 & 11 Flint & A &  6:55 & 0.178 & 14.7 &           &      \\
1311, Jul 16 &  1 Reed  & T & 10:33 & 0.923 & 68.2 &           &      \\
1313, Nov 18 &  3 House & T & 17:40 & 0.436 &$-$4.9&           &      \\
1314, Nov 08 &  4 Rabbit& T &  5:50 & 0.563 &$-$5.4&           &      \\
1318, Aug 26 &  8 Rabbit& H & 11:22 & 0.822 & 75.4 &           &      \\
1325, Apr 13 &  2 House & T & 10:27 & 0.987 & 66.9 & bad weather? 
                        & Fig.\ \ref{fig:se1325}; \cite{espenak}; E.K. \\
1326, Sep 26 &  3 Rabbit& A & 17:45 & 0.723 &$-$0.6&           &      \\
1329, Jan 30 &  6 House & A & 15:45 & 0.718 & 28.1 &           &      \\
1332, Nov 18 &  9 Flint & T & 17:43 & 0.357 &$-$5.4&           &      \\
1333, May 14 & 10 House & A &  5:47 & 0.430 &  4.7 &           &      \\
1337, Aug 26 &  1 House & T &  8:00 & 0.383 & 31.0 &           &      \\
1340, Jun 25 &  4 Flint & A & 12:47 & 0.801 & 79.1 &           &      \\
1340, Dec 19 &  4 Flint & T & 10:23 & 0.605 & 40.9 &           &      \\
1348, Jul 26 & 12 Flint & H & 17:33 & 0.589 & 12.9 &           &      \\
1349, Dec 10 & 13 House & T &  8:57 & 0.240 & 28.4 &           &      \\
1351, May 25 &  2 Reed  & H & 15:40 & 0.475 & 38.0 &           &      \\
1355, Mrc 14 &  6 Reed  & A &  8:35 & 0.851 & 34.7 & bad weather? &   \\
1357, Jul 17 &  8 House & A & 15:34 & 0.292 & 41.1 &           &      \\
1359, Dec 20 & 10 Reed  & P & 10:46 & 0.156 & 43.4 &           &      \\
1362, Oct 18 & 13 Rabbit& A &  5:35 & 0.419 &$-$6.4&           &      \\
1363, Oct 07 &  1 Reed  & P & 13:57 & 0.074 & 47.4 &           &      \\
1365, Aug 17 &  3 House & T &  8:40 & 0.677 & 40.8 &           &      \\
1368, Dec 10 &  6 Flint & H &  7:58 & 0.699 & 17.5 &           &      \\
1372, Sep 27 & 10 Flint & H &  8:22 & 0.729 & 57.7 &           &      \\
1376, Jan 21 &  1 Flint & T & 15:11 & 0.120 & 33.0 &           &      \\
1379, May 16 &  4 Reed  & T &  8:25 & 0.508 & 40.7 & (total in Phoenix/Aztlan) & \\
1380, Oct 28 &  5 Flint & A & 16:59 & 0.656 &  5.1 &           &      \\
1383, Mrc 04 &  8 Reed  & A*& 13:37 & 0.951 & 59.1 & bad weather? &   \\
1391, Sep 28 &  3 Reed  & T &  8:11 & 0.407 & 31.1 &           &      \\
1394, Jul 28 &  6 Rabbit& A &  6:07 & 0.632 &  5.8 &           &      \\
1395, Jan 21 &  7 Reed  & T & 14:31 & 0.754 & 40.1 &           &      \\
1397, May 26 &  9 House & T & 18:55 & 0.748 &$-$4.9&           &      \\
1398, May 16 & 10 Rabbit& T &  9:38 & 0.422 & 57.7 &           &      \\
1402, Aug 28 &  1 Rabbit& A & 14:53 & 0.898 & 45.9 & bad weather? &   \\
1404, Jan 12 &  2 Reed  & T & 13:04 & 0.441 & 48.7 &
                       Huichapan, 7v (p14) 
                       & $\;\;$\begin{rotate}{90}$\Lsh$\end{rotate} mistaken?
                         \cite{quirozennis_2016}; \cite{aveni_1999} \\
1405, Jun 26 &  4 House & H & 10:21 & 0.990 & 65.8 & (bad weather?) &  \\
1409, Apr 15 &  8 House & A &  5:49 & 0.468 &  2.0 &           &      \\
1410, Apr 04 &  9 Rabbit& P &  6:27 & 0.182 &  9.0 &           &      \\
1412, Aug 07 & 11 Flint & A & 15:43 & 0.415 & 35.1 &           &      \\
1416, Nov 19 &  2 Flint & A &  5:36 & 0.753 &$-$9.9&           &      \\
1417, Nov 08 &  3 House & P & 15:06 & 0.264 & 27.9 &           &      \\
1419, Sep 19 &  5 Reed  & T &  8:05 & 0.251 & 30.6 &           &      \\
1423, Jan 12 &  9 Reed  & H & 11:31 & 0.264 & 49.5 &           &      \\
1424, Jun 26 & 10 Flint & T &  6:09 & 0.344 &  8.6 &           &      \\
\end{tabular}
\end{table*}


\begin{table*}[t!]
%
%
\changefontsizes{9.5pt}
\centering
\begin{tabular}{l|r|c|r|c|r|p{3.8cm}|p{2.9cm}}
\hline
\rowcolor{grey20}
Date         & Aztec year & Type &\multicolumn{1}{c|}{\cellcolor{grey20}LT}
                                    & Magn. & Alt. & Codex     & Ref. \\
\hline
1426, Oct 30 & 12 Rabbit& H & 11:24 & 0.799 & 53.8 &
                       Vaticanus, Plate 119? 
                       & \cite{gallatin_1845}; \cite{milbrath_1995} \\
1427, Apr 26 & 13 Reed  & A & 15:31 & 0.111 & 39.1 &           &      \\
1430, Feb 22 &  3 Rabbit& T & 16:36 & 0.657 & 20.3 &           &      \\
1434, Nov 30 &  7 Rabbit& A & 18:26 & 0.945 &$-$1.3& bad weather? &   \\
1437, Apr 05 & 10 House & A*&  9:03 & 0.947 & 46.1 &           &      \\
1438, Mrc 25 & 11 Rabbit& A & 14:58 & 0.250 & 45.5 &
                       Telleriano, 31r; 
                       Vaticanus A, 76r 
                       & $\;\;$\begin{rotate}{90}$\Rsh$\end{rotate} record mistaken?
                         Fig.\ \ref{fig:eclipseglyphs}b; E.K. \\
1442, Jul 07 &  2 Rabbit& T & 16:57 & 0.442 & 22.3 &           &      \\
1444, May 17 &  4 Flint & H & 16:14 & 0.221 & 29.9 &           &      \\
1449, Feb 22 &  9 House & T & 16:40 & 0.081 & 19.5 &           &      \\
1451, Jun 28 & 11 Reed  & T & 17:10 & 0.384 & 19.4 &           &      \\
1452, Jun 17 & 12 Flint & T &  7:17 & 0.896 & 24.3 & bad weather? &   \\
1455, Apr 16 &  2 Reed  & A & 19:03 & 0.430 &$-$10.5&
                       Huichapan, p130(?) 
                       & $\;\;$\begin{rotate}{90}$\Lsh$\end{rotate} mistaken?
                         \cite{aveni_1999} \\
1456, Sep 29 &  3 Flint & A & 12:54 & 0.745 & 59.8 &
                       Huichapan, 19 (p38) & \cite{quirozennis_2016} \\ 
1457, Sep 18 &  4 House & A & 17:48 & 0.103 &  0.0 &           &      \\
1459, Jul 29 &  6 Reed  & T &  6:50 & 0.378 & 15.7 &           &      \\
1466, Sep 09 & 13 Rabbit& A & 10:59 & 0.513 & 67.9 &           &      \\
1467, Aug 29 &  1 Reed  & A & 10:45 & 0.239 & 68.0 &           &      \\
1470, Jun 28 &  4 Rabbit& T & 17:38 & 0.267 & 13.1 &           &      \\
1477, Feb 13 & 11 House & H & 13:59 & 0.891 & 51.1 &
                       Mendoza, 10r; Telleriano 
                       & Fig.\ \ref{fig:eclipseglyphs}c+d + \ref{fig:axayacatl};
                         \cite{berdan-anawalt}; \cite{deleon-ball} \\
1480, Dec 01 &  1 Flint & H & 14:52 & 0.535 & 29.5 &
                          Aubin, p37 
                        & $\;\;$\begin{rotate}{90}$\Lsh$\end{rotate} mistaken?
                          \cite{aveni_1999} \\
1481, May 28 &  2 House & A &  8:48 & 0.809 & 45.7 & 
                       Chimalpop.; Chimalpahin 
                       & Fig.\ \ref{fig:axayacatl};
                         \cite{chimalpopoca}; \cite{aveni_1999} \\
1484, Mrc 26 &  5 Flint & T & 16:55 & 0.621 & 18.0 &           &      \\
1485, Mrc 16 &  6 House & T &  6:39 & 0.252 &  8.2 &           &      \\
1488, Jan 13 &  9 Flint & A & 17:33 & 0.233 &  2.6 &           &      \\
1489, Jan 01 & 10 House & A & 18:14 & 0.514 &$-$7.9&           &      \\
1491, May 08 & 12 Reed  & A &  4:52 & 0.812 &$-$7.9& Chimalpopoca, p118
                        & misdated from 1490? \\
1492, Apr 26 & 13 Flint & A & 11:17 & 0.528 & 80.1 &
                        Chimalpopoca; Mexicanus 
                        & \cite{chimalpopoca}; \cite{aveni_1999} \\
1494, Mrc 07 &  2 Rabbit& T &  6:03 & 0.272 &$-$2.1& Chimalpopoca, p118 & misdated from 1493? \\
1496, Aug 08 &  4 Flint & T & 15:05 & 0.957 & 46.2 &     
                       Chimalpahin, p119; 
                       Chimalpopoca, p119; 
                       Vaticanus A, Plate 124?; 
                       Telleriano, 40v 
                       & Fig.\ \ref{fig:eclipseglyphs}e + \ref{fig:axayacatl};
                         \cite{aveni_1999}; \cite{deleon-ball} \\
1497, Dec 23 &  5 House & P & 10:54 & 0.101 & 46.1 &           &      \\
1499, Jun 08 &  7 Reed  & A & 18:24 & 0.625 &  2.3 &
                       Torquemada, p192 & \cite{aveni_1999} \\
1503, Mrc 27 & 11 Reed  & T & 17:30 & 0.139 &  9.9 & 
                       Chimalpopoca, p120 
                       & $\;\;$\begin{rotate}{90}$\Lsh$\end{rotate} mistaken?
                         \cite{chimalpopoca} \\
1504, Mrc 16 & 12 Flint & P &  5:40 & 0.468 &$-$5.4&
                       Torquemada; 
                       Telleriano 42r; 
                       Mexicanus, Plate 79(?) 
                       &\cite{aveni_1999} \\
1505, Jul 30 & 13 House & T & 15:58 & 0.007 & 41.6 &
                       Mexicanus; Aubin, p40 
                       & mistaken $\pm$1? \cite{diel_2018}; \cite{aveni_1999} \\
1506, Jul 20 &  1 Rabbit& T &  5:18 & 0.671 &$-$3.9&           &      \\
1508, Jan 02 &  2 Reed  & A &  7:55 & 0.418 & 15.3 &
                       Huichapan, p30; 
                       Chimalpopoca; 
                       Telleriano 42r;
                       Torquemada, p210; 
                       Vaticanus A, 84v;
                       Bilimek/Borgia 
                       & Fig.\ \ref{fig:huichapan} + \ref{fig:bilimek};
                         \cite{quirozennis_2016};
                         \cite{chimalpopoca};
                         \cite{aveni_1999};
                         \cite{taube_1993}; 
                         \cite{milbrath_1997} \\
1510, Nov 01 &  5 Rabbit& A & 12:51 & 0.376 & 50.2 &
                       Telleriano, p42v; Vaticanus A, 87v; Mendoza, 16r? 
                       & \cite{milbrath_1997}; \cite{gallatin_1845}; E.K. \\
1511, Oct 21 &  6 Reed  & A & 16:05 & 0.277 & 17.8 &           &      \\
1516, Dec 23 & 11 Flint & T &  5:45 & 0.690 &$-$11.9&
                        Chimalpahin, p121 & \cite{aveni_1999} \\
1520, Oct 11 &  2 Flint & A &  7:51 & 0.058 & 24.9 &           &      \\
\cdashline{1-8}[0.5pt/5pt]     
1521, Sep 30 &  3 House & A &  7:56 & 0.713 & 27.5 &           &      \\
1524, Jul 30 &  6 Flint & T & 15:41 & 0.686 & 38.5 & Mexicanus, p56;
                          Aubin, p62 
                        & Fig.\ \ref{fig:eclipseglyphs}f;
                          \cite{diel_2018}, \cite{aveni_1999} \\
1525, Jan 23 &  7 House & A &  7:32 & 0.239 & 11.7 &           &      \\
1531, Mrc 18 & 13 Reed  & H & 14:00 & 0.542 & 57.5 &
                       Telleriano, 44r & \cite{deleon-ball} \\
1535, Jan 03 &  4 Reed  & T & 17:46 & 0.717 &$-$1.6 &          &      \\
1538, Apr 28 &  7 Rabbit& T & 16:05 & 0.252 & 31.3 &           &      \\
1539, Apr 18 &  8 Reed  & T &  6:27 & 0.115 & 11.3 &           &      \\
1542, Feb 14 & 11 Rabbit& A & 16:50 & 0.323 & 16.4 &           &      \\
1543, Jul 31 & 12 Reed  & A &  9:28 & 0.460 & 52.6 & (Telleriano, 46r) & (E.K.) \\
1546, May 29 &  2 Rabbit& H &  7:10 & 0.738 & 23.4 &           &      \\
%
%
\end{tabular}
\end{table*}


The Aztecs did not record calendric days on which eclipses
occurred, but the year can be in error by $\pm$1 though.
Table \ref{tab:eclipses} lists all eclipses that would have
principally been visible from the capital Tenochtitlan in the 14th
and 15th century until well beyond the conquest.
Most of them were of low magnitude, and they usually escape
attention of an unprepared.

Most records belong to the last 52-year cycle that preceded the
Spanish invasion.
Conspicuous obscurations from the 14th century are missing, while
other phenomena like severe droughts and locusts plagues were put
to record.
For example, \textit{Chimalpahin} and \textit{Mexicanus} tell about
a drought in 1332 and 1354--55, respectively.
Dendrochronology verifies a severe drought in the absolute years
1332 to 1335 \cite{therrell_2004},
however, this period is not mentioned among the so-called
``megadroughts'' \cite{stahle_2011}.
On balance, we can infer that a good part of the chronicles may be
reliable.

\begin{figure}[t]
\centering
\includegraphics[width=\linewidth]{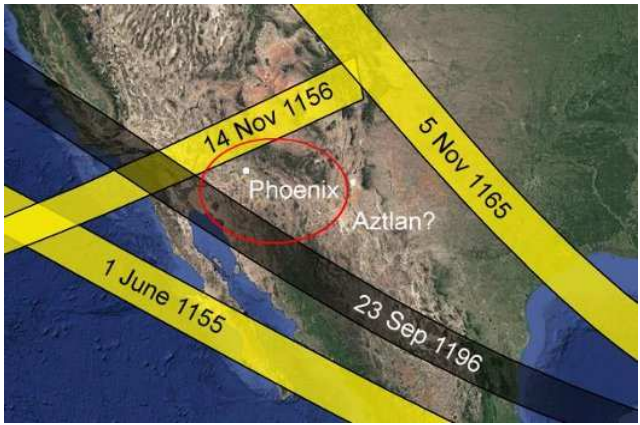}
\caption{Three annular eclipses in the mid-12th century in the
    vicinity of Aztlan.}
\label{fig:aztlan}
\vspace{-1ex}
\end{figure}

\paragraph{1155--1165, 1196:}
In the region what is meant to be ``Aztlan'' in the northwest of
Mexico, there were three eclipses of magnitude $>$0.8 within 10
years (Fig.\ \ref{fig:aztlan}).
The years are not far in time from the putative start of migration.
It will be extremely speculative to claim this series of eclipses
to have an effect on the people's decision to leave their homeland.
In history of mankind there is no paradigm that someone left his
hearth and home behind because of an eclipse.
One would rather look for other causes for migration like economic
and social reasons.
There is evidence from dendrochronology about an extreme drought
in the western North America lasting from 1149 to 1167
\cite{stahle_2011}.

Another striking event took place in 1196.
If someone gives credit to the Aztec chronology, the migrants had
already left their homeland.
The comparison of three codices (\textit{Boturini},
\textit{Azcatitlan}, and \textit{Aubin}) worked out essentially
identical itineraries for the migration \cite{rajagopalan}.
According to the timeline in \textit{Azcatitlan}, the capital was
founded in 1354 (instead of 1325, see below).

\begin{figure}[t]
\centering
\includegraphics[width=\linewidth]{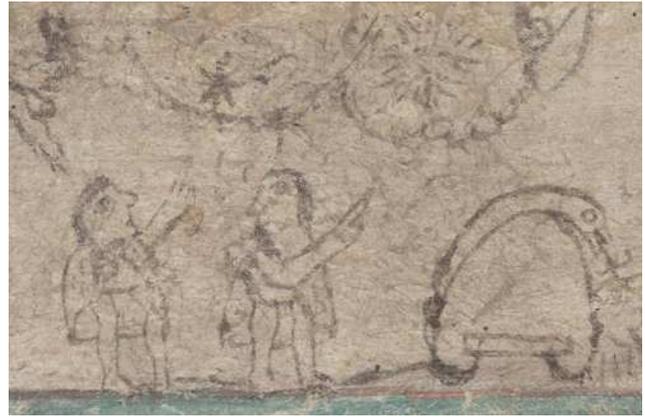}
\caption{Two persons pointing at a celestial event:
    a star and the sun, their order probably reversed,
    in $\approx$1199.}
\label{fig:comet1199}
\end{figure}

\enlargethispage{1ex}

\paragraph{1199:}
A celestial event is indicated in the \textit{Mexicanus}
(Fig.\ \ref{fig:comet1199}).
Two persons are pointing at a star and the sun.
The picture is placed at 5 Rabbit (1198) which is assumed the 31st
year of migration.
Anthony Aveni interprets the allusion to the sun as an eclipse in
1198 \cite{aveni_1999}.
It would have occurred on a winter's day at sunset.

The collection of historical comets by Donald Yeomans holds an
entry from China and Korea \cite{yeomans_1991}.
A comet appeared for three weeks from 16 August to 6 September 1199
moving in the northern sky from Hercules through Draco to Ursa
Major.
The parameters of the comet suggest its close passage by the earth
at a distance $d \approx$ 0.06 astronomical units (9 million km).
We assume an error in the \textit{Mexicanus} by 1 year as well as
the sequence interchanged:
the eclipse of a somewhat larger magnitude occurred on 24 July 1199
and prior to the appearance of the comet.

\paragraph{1301, 1303, or 1311:}
A monstrous animal devouring the sun is shown in the
\textit{Azcatitlan} while the group of people was continuing its
migration from a hill Yohualtecatl near Guadalupe
(Fig.\ \ref{fig:eclipseglyphs}a).
The event corresponds to an eclipse, indeed, and increases the
veracity of the historical record.
On the other side, the location names are badly identified and
not mentioned in other codices.
The timeline does not provide an exact year, but 1301 is favoured
by us, though the second eclipse in 1303 must have been conspicuous,
too, for its magnitude reaching 0.83 precisely at sunset.
The sun at a low altitude has the advantage of glancing into its
dimmer light reducing eye damage.
The year 1311 cannot be ruled out, either.

\paragraph{1325:} \label{ch:tenochtitlan}
The foundation of Tenochtitlan is wrapped in tales.
The date is specified to 14th March 1325, the day after spring
equinox in the year after the arrival of the migrants from Aztlan
at Lake Texcoco.
That is not to say that the foundation \emph{really} happened in
this year nor on that precise day.
The date just equals the beginning of the second solar year of
the 52-year cycle.
The choice strongly suggests a fabrication by the Aztec chroniclers
to adjust a desired progress of events.

\begin{figure}[t]
\centering
\includegraphics[width=\linewidth]{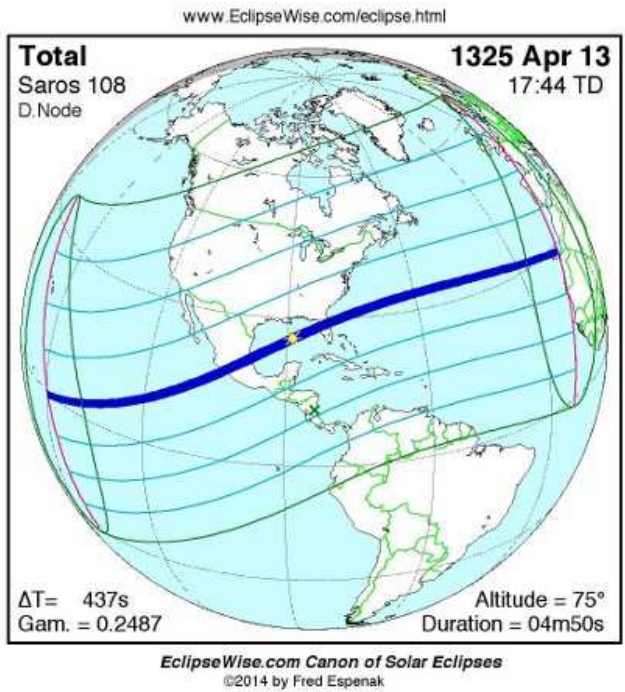}
\caption{The total solar eclipse of 13 April 1325 flashed by the
    Aztec capital Tenochtitlan \cite{espenak}.}
\label{fig:se1325}
\end{figure}

Modern historians tend to expand this date to involve a
quasi-astrological relationship \cite{aveni_1999}:
The day of 14 March 1325 was New Moon.
Saturn, Jupiter, and Mars were approaching each other and gave an
impressive clustering in the sky.
Since the three planets are the slowest wanderers and meet least
often, the Maya associated with this configuration a creation myth,
though no such context is known for the Aztecs.
The planets attained their narrowest separation on 11 April 1325.
Two days later a total eclipse occurred (Fig.\ \ref{fig:se1325}).
The zone of totality missed the city of Tenochtitlan by 35 km, but
an obscuration of 99\% will be impressive enough.

A record about this eclipse is not known, and there seem to be no
myths about it.
Usually, the sudden loss of daylight leaves many traces worldwide,
and they are converted into legends that would go so far as to
concern the rise and fall of cultures.
From the old continents there are examples of legends encompassing
the foundation of Rome, the construction of Amarna, the sack of
Babylon, or the end of Ugarit \cite{finsternisbuch, khalisi-egypt}.
The Indian tribes in the United States farther north considered
eclipses as a ``moment of renewal'', and the idea still holds
today.
For Tenochtitlan there \emph{does} exist an appropriate eclipse of
large magnitude, but no-one seemed to care.
The coincidence of astrological signs in the sky cries out for
wrapping the establishment of the nation into a legend.
From its non-existence we draw the conclusion that either bad
weather prevented observation or the alleged date of the foundation
is wrong.

\paragraph{1404:}
The observation of the partial eclipse of such low a magnitude
(mag = 0.44) at noontime is anything but self-evident.
We attribute it to unknown advantageous conditions, as it caught
someone's attention by chance:
a slight cloudiness unharmful for the eye, or the lowered
altitude of the sun at winter's time.
The event was followed by an almost total eclipse of the moon
15 days later.
Much larger obscurations of the sun occurred in summer of 1402 and
1405, respectively.
It is possible that the writer was mistaken at counting to one
side or the other.

\paragraph{1426:}
Plate 5 of the \textit{Telleriano} says that ``the earth became
eclipsed'' \cite{gallatin_1845}.
We cannot verify whether the account refers to the solar eclipse
of 30 October.
The magnitude of 0.8 is on the verge of an accidental
detectability.

Another statement concerns that this eclipse ``presaged'' the
death of the king Chimalpopoca \cite{milbrath_1995}.
Most historians put his death a quarter of a year later, at least,
in 1427.
According to the \textit{Codex Chimalpopoca} that king was murdered
in 1428, and there is no mention of an eclipse.
Also, \textit{Mendoza} does not tell anything about it.
The time gap seems too large to draw a direct connection with
the eclipse.
In general, we cannot confirm the concept of relating such events
to the death of rulers, because there is no single example.
No codex gives a clue to such an association.
It seems an idea from the old world rather than of the cultures in
Mesoamerica.

\paragraph{1437:}
A sun symbol appears on page 76v of the \textit{Vaticanus A}
with a tie to a warrior that himself is close to the plaque of
either 12 Reed (1439) or 11 Rabbit (1438), see Figure
\ref{fig:eclipseglyphs}b.
A large obscuration happened in 1437, and we take the liberty of
assigning the symbol to this year.

\begin{figure}[t]
\centering
\includegraphics[width=\linewidth]{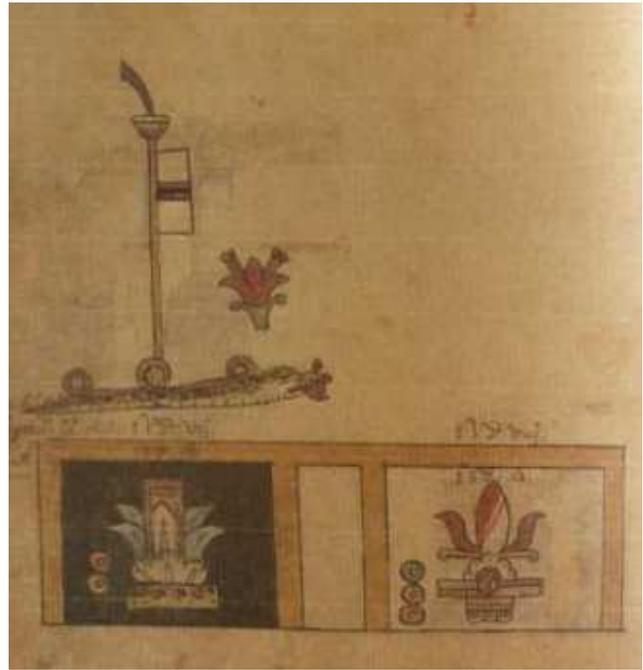}
\caption{Year sign for 1508 in the \textit{Huichapan}
    \cite{quirozennis_2016}.}
\label{fig:huichapan}
\end{figure}

\paragraph{1456:}
The \textit{Huichapan Codex} mentions three eclipses:
1404, 1456, and 1508.
They are said to have occurred in the years of the New Fire
ceremony, albeit not on the exact days of the festival.
All magnitudes were small to medium, and it is absolutely not
clear whether or not the following interpretation of the glyphs
is correct:
Above the sign ``2 Reed'', which is drawn on a black background for
all three years, the last eclipse of 1508 shows a snake
(Fig.\ \ref{fig:huichapan}).
It carries three circles on its body and a flag standing on the
center of it (second circle).
An idea by Rossana Quiroz Ennis for its meaning concerns a possible
52-year-cycle of eclipses \cite{quirozennis_2016}.

Mathematics involves only the counting of days between the three
eclipses yielding two intervals of slightly different length:
19,254 and 18,722 days, respectively.
The asymmetric intervals equate to an integer of synodic months
and half-integer of draconitic months, as will be required for an
eclipse cycle.
Especially, those 18,722 days of the second interval are equivalent
to a ``Thix period'' taken twice.
The Thix is an interval of 9360 days plus one, known to the Maya,
as they experimented with Venus- and eclipse-cycles
\cite{smiley_1973}.
It is composed of 36$\times$ the sacral 260-day-calendar,
26$\times$ the 360-day-year, and coming close to 16$\times$ the
synodic period of Venus of $\approx$584 days (actually, 1 day more).
Although such a period is deprived of a physical basis, the Maya
(and later the Aztecs) \emph{could} have assumed such a cycle.
Being unaware of a better solution, they arrived at a good guess,
indeed.

However, when the next eclipse will be due after those 26 solar
years, the observer is displaced by almost 5 hours in regard to
the previous scene.
It is a matter of chance that we find this sequence of the Thix
series in our list.
For example, the intermediate event of 1430, which would account
for the next incident after 1404, might have escaped attention
because of any reason and, therefore, not recorded.

We want to emphasise that the interpretation of this glyph in
favour of the 52-year-cycle is appealing, but as unsteady as any
other suggestions, because the Aztecs did not compute cycles.
For the discovery of a long-term cycle, systematic records are
mandatory.
The task usually commences with much shorter periods, tentatively
within a generation, before being expanded to longer ones.
An Aztec mathematician --- if any existed --- might have
\emph{surmised} such a cycle, but we cannot confirm that he was
lucky in identifying this special occasion in his chronicles.
Mathematics was poorly conceived, or, to put it more gently:
we do not know about its level.
On one hand, we lack of genuine manuscripts from the pre-hispanic
era, and, on the other hand, the Spanish missionaries did not
provide helpful information about that.
Probably they were themselves hardly familiar with mathematics and
did not understand what they were told.
In this special case we just meet the interval of 52 years because
of three pillars pegged (not very well) to the year designation
``2 Reed''.
The suggestion for this cycle is a nice ansatz but not supported
by crude facts.

\paragraph{1470ies:}
Anthony Aveni cites the Spanish friar and chronicler Juan de
Torquemada (1562?--1624) as a source for three accounts linked to
the years 1473, 1475, and 1476 \cite{aveni_1999}.
Each is accompanied by a bad omen:
the death of a king in a neighbouring city,
a battle as well as an earthquake,
and the wounding of the Aztec ruler in another battle,
respectively.
However, there were no solar eclipses in any of these years to be
observed in Mexico.
Two eclipses of the moon did take place (1473 and 1475),
but it seems futile to invent a hypothesis to please the three
accounts.
Torquemada is the only writer known having implanted these
omen-related eclipses into history (see ``1499'' below).
Unfortunately, no English translation of his work exists,
and we cannot trace back how he got the information about it.

In a like manner the native writer Chimalpahin (1579--1660) states
that eclipses occurred in 1476, 1478, and 1479.
For the event of 1478, he writes that it was total such that
``\dots stars were visible on the day 1 Movement''.
Aveni believes that Chimalpahin could have used a non-Aztec
calendar or misread a pictorial source, but he does not provide
a suggestion for a correct reading to understand the error.
What other calender systems were in use besides the Aztec?

\textit{Telleriano} on page 37r connects the glyph of a partially
covered sun with 1476 (Fig.\ \ref{fig:eclipseglyphs}d),
while \textit{Vaticanus A} (Codex \#3738), which seems to be in
large parts a copy of the \textit{Telleriano}, shows the same
glyph close to a warrior in 1478 or 1479.
Since there were no eclipses in Mexico in any of these years, all
entries should be considered a fault for 1477.

\begin{figure}[t]
\centering
\includegraphics[width=\linewidth]{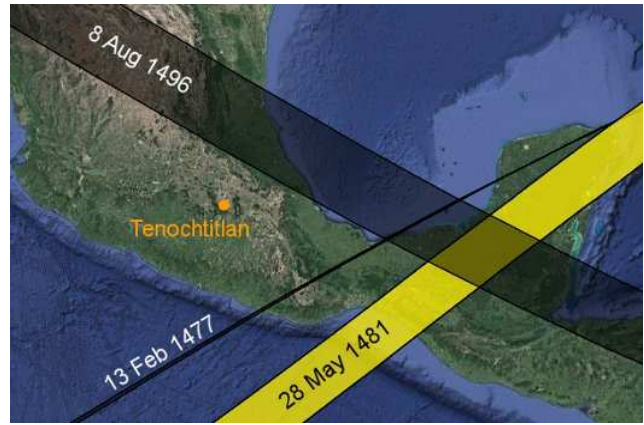}
\caption{Two total eclipses (grey) and an annular eclipse (yellow)
    in 1477, 1481, and 1496.}
\label{fig:axayacatl}
\end{figure}

\paragraph{1477:}
The only notable event of the 1470ies happened on 13 February 1477
(Fig.\ \ref{fig:axayacatl}).
The \textit{Codex Mendoza} shows a cropped sun next to the symbol
for a defeated town (Fig.\ \ref{fig:eclipseglyphs}c).
The king Axayacatl conquered 37 towns between 1469 and 1481,
one of which was Tetenanco as the ninth \cite{berdan-anawalt}.
The eclipse of 1477 had the highest magnitude (mag = 0.89) during
the entire reign of that king.
We assign the \textit{Mendoza} glyph to this year.

\paragraph{1480:}
In the \textit{Codex Aubin} the sun symbol is placed at 1479 and
alludes to the ruler Axayacatl who died in the subsequent year.
The ruler died in 1481 (see below), so either the timeline is
short by one year, or the writer skipped a year.
It is unclear whether he really meant the small eclipse
(mag = 0.53) of 1480, or he is mistaken by 2 years up or down:
both 1477 and 1481 would fit.
In all, there is much confusion concerning the whole decade.

\paragraph{1481:}
Mexican historians who started to write after the conquest mention
two great eclipses of the sun that happened in the interval of five
years and of the events which preceded them \cite{gallatin_1845}.
The first eclipse occurred after the victory of the king Axayacatl,
and the other after the death of the same.
With this information at hand we can fix the two eclipses to 1477
and 1481 implying an ``inclusive counting''
(Fig.\ \ref{fig:axayacatl}).
Both events had the same year designation except for the numerical
character.
The native naturalist Chimalpahin asserts that Axayacatl died before
7th June 1481, and his successor was installed the second day after.

\paragraph{1491:}
\textit{Chimalpopoca} tells about eclipses of the sun in 1490 and
1492 and that stars appeared \cite{chimalpopoca}.
Actually, the first eclipse took place over the South Pole in
Antarctica.
The writer must be mistaken by one year to one side or the other:
\begin{compactitem}
\item On 1 January 1489, an observer would have principally been
able to watch the very beginning of an eclipse during sunset.
After about fifteen minutes, the sun sank below the mathematical
horizon with a magnitude of 0.27.
\item In 1491, the sun would rise partially eclipsed on 8 May with
a magnitude of 0.38 and decreasing.
Within 25 minutes after its appearance above the horizon, the
disk would be cleared.
\end{compactitem}

The qualitative chances for visibility are almost the same for
both eclipses.
Since the date of 8 May 1491 gives a smaller distance to the
historical context in Chimalpopoca's chronology, we prefer this one.
But there is no justification for the statement on stars visible.
An eclipse in 1492 is correct, but the entry itself can be in error.

\paragraph{1494:}
\textit{Chimalpopoca} records an eclipse for 1493, but it should
be shifted to the following year.
Similar to the case of 1491 above, it happened immediately at
sunrise with a magnitude of 0.26 and diminishing.
The additional statement ``stars appeared'' is wrong, either.

\paragraph{1496:}
Inspite of the impressively high magnitude of this spectacle, it
is missing in several important manuscripts
(Fig.\ \ref{fig:axayacatl}).
\textit{Telleriano} shows a new style of an eclipse image,
differing from the glyph on the previous page
(Fig.\ \ref{fig:eclipseglyphs}e).
Having stars and a crescent moon it goes with an European style
rather than with the Aztec style \cite{deleon-ball}.
Two pages further on, the classic Aztec glyph is used for 1508
again.

According to Anthony Aveni, the \textit{Codex Chimalpahin} speaks
of a ``\dots complete eclipse of the sun, so that it was as dark
as in the deepest night, and the stars were seen with complete
clarity'' \cite{aveni_1999}.
Again, this is an exaggeration, for only Venus and Mercury would
have shown up, and probably Sirius as the brightest star.
Mars was close by with a stellar magnitude of +2$^{\rm m}$ only.
It is doubtful that anybody took notice of these particulars
and just called them ``stars'' altogether.
Moreover, Aveni states in a note that there was a minor eclipse
on 3 January 1497, but we cannot confirm this.
He probably meant the event on 2 February of that year, but it
was only seen from Antarctica and the adjacent oceans.

\paragraph{1499:}
The report is given by Juan de Torquemada only and embedded into
a sense of divination \cite{aveni_1999}.
It was to ``\dots\ announce an inundation and a great famine''.
The massive flood is mentioned in every Aztec codex implying that
it turned into one of the greatest disasters of the Aztec history.
It lasted for three or four years till 1502.
The \textit{Codex Chimalpopoca} adds that the earth shook four
times in 1499 \cite{chimalpopoca}.
Next year waters spread out everywhere reaching other cities.
The codices \textit{Mexicanus} and \textit{Aubin} depict streams
of water and a man carrying a stone, respectively.
According to \textit{Codex Duran}, Tenochtitlan had to be rebuilt
after the flood \cite{diel_2018}.
Since crops were destroyed, a famine followed.

The eclipse itself did not cause that bunch of disasters, of
course, but we can reflect on why only Torquemada, who lived a
century later, mentions it in connection with the other affairs:
the eclipse slid into the background upon severe problems to be
managed;
or he contrived the fateful connection of natural phenomena
announced by God from his religious education;
or the eclipse was not observed at all, and he amended it
fictitiously to enhance the effect of his version of the story.
Acts like the latter were quite common in history and they still
are, see \cite{finsternisbuch} for various examples.
The ``identification'' of the eclipse is wrong then.

\paragraph{1504:}
\textit{Chimalpopoca} mentions an eclipse in 1503.
A partial obscuration took place, indeed, on 27 March of that year
in the late afternoon before sunset (mag = 0.14).
However, the native writer gives in his chronicle one numeral less
for the year than it should be (compare the entries ``1491'' and
``1494'' above).
We proceed from this latter assumption and believe that the event
should be placed correctly in 1504.
On 16 March 1504 at 7:04 a.m., the sun rose above the horizon with
a magnitude of 0.34 and decreasing.
This would be the third small obscuration at the early hours of
sunrise.

\begin{figure*}[t]
\includegraphics[width=\linewidth]{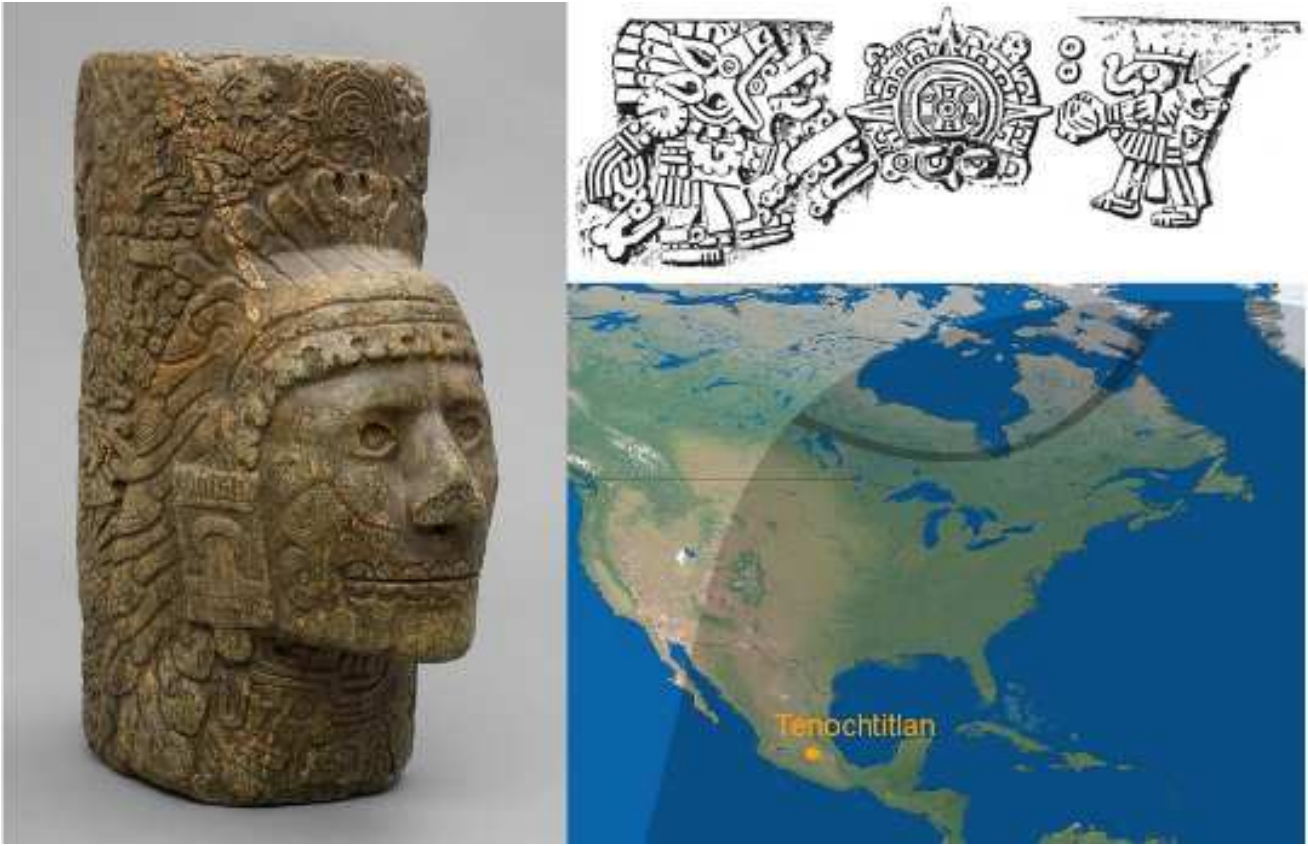}
\caption{The Bilimek Vessel with an eclipse sign (above the
    skeletised head), which is menaced by two Tzitzimime.}
\label{fig:bilimek}
\end{figure*}

\paragraph{1505:}
The eclipse was not visible from central Mexico.
Again, it remains a mystery why the artists of some codices
recorded it.
They related it with those Tzitzimime, the ``eclipse demons'',
that are said to descend to earth during totality, but there was
no totality far and wide.
A misdating with one prior or later year is unlikely, for both
proximate years comprised weak partial eclipses only.
For the previous year, see ``1504'' above.
One year later, on 20 July 1506 at 6:41 a.m., the sun rose obscured
at half (mag = 0.56) and decreasing, too.

The eclipse of 1505 was to occur shortly before the completion of
the 52-year cycle.
Perhaps the post-conquest missionaries were targeted at anxiety
among the Aztecs and designed the Tzitzimime in connection with
the cataclysmic events at the end of the 52-year count?
Also, a misinterpretation of the New Fire ceremony cannot be
ruled out.

\paragraph{1508:}
This eclipse has been discussed in literature several times.
\textit{Telleriano} connects the sun glyph with a warrior, the
completion of a temple, an earthquake, the drowning of 1,800 men
in a river, and the New Fire ceremony.

Another symbol for this event can be found on the Bilimek Vessel
(Fig.\ \ref{fig:bilimek}).
The vessel, stored in the Museum f\"ur V\"olkerkunde in
Vienna/Austria, is covered with iconographic reliefs.
Of particular interest is a partly covered sun above the head.
The sign differs slightly from other illustrations in that the
lower portion of the sun is curved, much like the edge of the
moon during a solar eclipse.
On each side there are figures, identified as the ``eclipse demons'',
Tzitzimime, menacing the sun with stones and wooden staffs from
both sides \cite{taube_1993}.
However, the magnitude of this eclipse was far from having any
effect on sunlight.
Since the Tzitzimime are creatures of darkness, their appearance
is highly questionable in this context.

Other icons on the vessel are accompanied by circular elements.
The meaning of the small circles is still unclear.
Eduard Seler interpreted them as numerical coefficients of
calendrical dates,
Karl Taube suggests certain stars,
and they can also be ornaments without any meaning.

The vessel holds at least three and probably four items appearing
in the codices \textit{Borgia} and \textit{Vaticanus B}.
The sign of an earthquake on the underside of the vessel lets
Susan Milbrath assume that it took place in the same year as the
solar eclipse \cite{milbrath_1995}.
In combination with the Tzitzimime-argument she concludes that
the vessel was manufactured shortly after the events in 1508.
However, earthquakes happen in Mexico so often that their
coincidence with an eclipse becomes of no use for dating purposes.
The years 1480, 1489, 1496, and 1499 harboured this pair of
catastrophes as well, while the combination with the New Fire
ceremony is true for 1455/1456.
The sole argument in favour of the year 1508 is the shape of the
sun itself:
the moon covered the lower part of the disk agreeing with the
approximate view from central Mexico.

\paragraph{1510:}
\textit{Codex Mendoza} shows the glyph of a solar eclipse next to
two conquered towns during the reign of Moctezuma II (1502--1520).
The years are not given, so we cannot assign an eclipse to the
glyphs.
The years 1508 and 1510 had the largest magnitudes among the
options.
Possibly the two towns were defeated shortly after another in the
same year.

\paragraph{1516:}
Aveni rejects the eclipse entry by Chimalpahin because of his
statement that ``no eclipses were visible in that year''
\cite{aveni_1999}.
Basically, this is true, if we neglect the diminishing obscuration
during sunrise at 7:16 a.m.
When the sun climbed the horizon, it exhibited a tiny bump
corresponding to a magnitude of 0.02 only.
It would be very remarkable, if such small an irregularity on a
perfectly circular disk would have caught any attention.

\paragraph{1521:}
This eclipse would be interesting in regard to its date.
It occurred one month after devastating Tenochtitlan (21 August
1521) and the massacre of the Aztec people by Hernan Cortes.
Actually, there seems to be no record for the eclipse.
A mistake with the eclipse of 1524 is possible, but it will be
difficult to explain the discrepancy of more than 1 year.
It is a remarkable trait of all codices that the accuracy of
the accounts on natural phenomena rarely exceeds $\pm$1 year.

\paragraph{1524:}
\textit{Codex Aubin} pictures the sun accompanied by ``8 Oct'' for
the year name 5 Reed (1523), but there was no eclipse in this year.
The \textit{Mexicanus} also shows a half-darkened sun
(Fig.\ \ref{fig:eclipseglyphs}f).
That day was usual New Moon, and the symbol of the sun does not
make sense in this context.
Maybe someone tried to predict an eclipse and failed?
The eclipse would fit better next year (6 Flint), but the day is
still wrong.
Alternatively, it may have a metaphorical meaning:
the collapse of the Aztec society led to an extinct light for the
indigenous culture.

\paragraph{1531, 1541, 1543:}
Perhaps the last eclipse glyph using the Aztec style is found in
the \textit{Telleriano} on page 44r and denotes the incident of
1531.
However, the last two images for the sun itself are on page 46r of
that codex and linked to the years 1541 and 1543, respectively.
An eclipse did take place on 31 July 1543, but the sun is drawn
uncovered and gives an impression of a torching summer responsible
for a drought.
Maybe the eclipse was not noticed on that day.


\section{Lunar Eclipses}

Accounts on eclipses of the moon were not kept.
The temporal darkening of the moon's face seem to be considered
``normal'', for it takes place almost every year.
\textit{Codex Telleriano} explicitly states in connection to the
solar eclipse of 1510 that ``\dots\ [the Aztecs] never took much
account of eclipses of the moon\dots '' \cite{aveni_1999}.
The mere existence of such a remark points to an observer who did
not experience that phenomenon for a longer period of time.
Suddenly he was reminded of it when it happened again, presumably
in connection with an event in 1511.
There was a period of 2,5 years between 1508 and 1511 when no
lunar eclipses were to be seen in Mexico.
In case of bad weather, the interval can be prolonged.

Albert Gallatin said that, when a lunar eclipse occurred, people
believed ``the sun would have eaten the moon'' \cite{gallatin_1845}.
It remains unclear whether the Aztecs really understood the cause
of eclipses, or this expression would be intermingled with modern
knowledge.
We plead for the latter.
Taking the view of an uninformed person, the phenomenon of a lunar
and solar eclipse presents itself very different in nature:
different in frequency, different in daytime, different in
length, different in the visual conditions.
It is not obvious at all to comprehend that an eclipse is nothing
more than an interplay of shadows caused by the bodies of earth or
moon.

The original sources do not permit an unequivocal view on the
reaction of the people when they unexpectedly saw the moon turning
red and disappearing.
Decapitated figures in the codices were presented as a proof that
the loss of heads would be equivalent to a loss of light in a
lunar eclipse \cite{milbrath_1997}.
We cannot join in this opinion because the pictographs are not
accompanied by calendric dates, especially concerning the moon.
If the hypothesis was true, the arrangement of the images would
coincide with celestial events, but this is not provided.
Moreover, after the moon will look restored, the incident will be
forgotten on a short timescale, while a decapitated being does
not recover.
This discrepancy impedes a parallel.

The same study lists twelve pairs of eclipses within a year between
1440 and 1511.
No accounts on actually \emph{observed} events are given, just
the astronomical dates.
The author assumes that all pairs were observed without constraints
to weather conditions.
``Disrupted'' pairs are not considered, either, i.e.\ those pairs
that span a year boundary.
It is much too speculative to construct a timeline for history
based on celestial opportunities only.
Astronomy is a strong assistant for historical research, but it
is not meant to deliver input data \textit{a priori} for a desired
story.


\section{A Note on Moctezuma II's Comet}

A widespread myth surrounds the Aztec king Moctezuma II who is
said to have sighted a comet that heralded the defeat of his
realm.
Various authors correlated the comet with diverse years like
1509, 1511, 1515--1519, or 1520/21. 

We tried to retrace the information, but we failed to find reports
on the comet apart from Aztec sources.
Neither the accounts from China nor Europe attest an observation
in those years, and this cannot be attributed to negligence.
For example, Johannes St\"offler (1452--1531), a German astronomer
operating on the verge of astrology, made foretellings on social
and political affairs.
He computed the course of planets, but he did not mention a comet,
though it would be easy for him to take advantage of such an
appearance.
Another humanist was Nicolaus Kratzer (1487--1550) who became an
astronomer at the court of Henry VIII in England --- also no
mention of a comet.
And the celebrated Nicolaus Copernicus (1473--1543) did not drop
any note to be in line with it.
The list of historical comets by Donald Yeomans \cite{yeomans_1991},
which is itself based on a catalogue by Ho Peng-Yoke and Ang
Tian-Se,
does not provide an expedient observation, either.
The closest appearances are entries for 1506 and for January
1520, respectively.
In 1520, a broom star was seen in China, but it must have been a
faint object visible for 29 days.
Neither the constellation nor a movement are mentioned. ---
So, where does the note on Moctezuma's comet originate from?


Book 8 of the \textit{Florentine Codex} by Bernardino de Sahagun
deals with kings, lords, and how they governed their reign.
There he lists eight bad omens for Moctezuma and repeats them in
Book 12 in more detail.
Two of them are celestial signs, the others must be terrestrial.
Omen \#1 is mentioned in a dozen codices, while \#4 deploys the
Spanish term ``cometa'', primarily introduced by Sahagun himself
\cite{pastrana}.
Omen \#1 is said to have appeared before dawn ten years before the
arrival of the Spaniards.
In Sahagun's words, the passage reads rather like an aurora of
impressive size, keeping in mind that some connotation might be
exaggerated.
More important is omen \#4, and it translates as follows
\cite{lockhart}:
\begin{quote}
The fourth omen was that while the sun was still out a comet fell,
in three parts.
It began off to the west and headed in the direction of the east,
looking as if it were sprinkling glowing coals.
It had a long tail, which reached a great distance.
When it was seen, there was a great outcry, like the sound of
rattles.
\end{quote}

\begin{figure}[t]
\includegraphics[width=\linewidth]{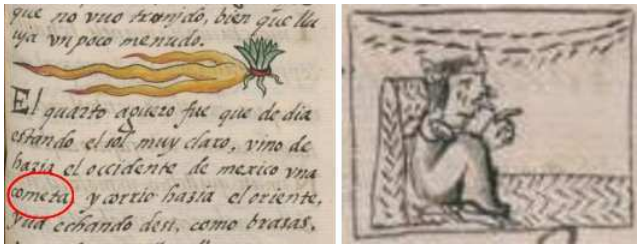}
\caption{Two images from Sahagun's \textit{Florentine Codex}
    on the description of omen \#4, from Books 8 and 12
    \cite{lockhart}.}
\label{fig:omen}
\end{figure}

The narrative resembles a fragmenting fireball rather than a comet,
but the friar probably did not know a better expression to use.
The incident happened at some time between 1515 and 1519, i.e.\
on the eve of the conquest, and caused great fear.
Both passages in Books 8 and 12 are accompanied by tiny images
(Fig.\ \ref{fig:omen}).
In any case, Sahagun was no eyewitness, and the small sketches
must be the result of his own imagination upon an oral interview.

\begin{figure}[t]
\includegraphics[width=\linewidth]{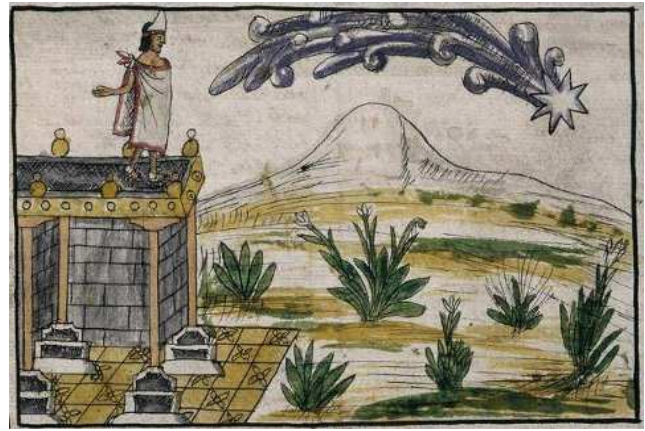}
\caption{Image from Diego Duran's \textit{History} \cite{duran},
    folio 182.}
\label{fig:montezuma}
\end{figure}

Another Spanish friar, who sailed as a child, together with his
parents, to the newly discovered colonies in Mexico was Diego Duran
(1537?--1588?).
He got to know about the indigenous culture, learnt their language,
and also tried to keep their tradition for the record.
His most important work is a book on the history of the Indians
composed at roughly the same time that Sahagun was composing his
\textit{Florentine Codex}, approximately 1574--81 \cite{duran}.
Duran had access to a number of pre-conquest pictorial manuscripts,
now lost, while Sahagun relied primarily on native informants.
There is no evidence that the two writers ever met, and their
writings seem unrelated to each other.
Duran speaks of an ``object at night'' in his book, and he included
that picture that was to become famous (Fig.\ \ref{fig:montezuma}).

Finally, the word ``comet'' is subsequently mentioned in two more
works dealing with those eight omens \cite{pastrana}.
First, in a manuscript by the Jesuit Juan de Tovar (1546?--1626?),
called the \textit{Tovar Codex} or \textit{Ramirez Codex}, for
the latter re-discovered it in 1856.
The completion of the manuscript is estimated to 1579, and
perhaps it is based on Duran's history book.
The second work is by Diego Mu\~{n}oz Camargo (1529?--1599), who
was born in Mexico as a son of an Indian mother.
He grew up bi-lingually, and this enabled him to become one of the
first Spanish-language chroniclers hearkening back to original
information of the natives.
Mu\~{n}oz Camargo also repeats the omens and writes that the
Lord of his hometown, when he was born upon the arrival of the
Spaniards, got his name after a ``great, horrific comet with a
great tail'' that was seen in the sky spewing smoke.
All descriptions about that phenomenon turn out very short,
vague, and can be misleading.

The \textit{Telleriano} depicts several geophysical events on
folio 42r covering the years 1507--1509.
The words describe a sort of \dots\
\begin{quote}
\dots\ \textit{mexpanitli} or ``cloud banner'' (cloud of smoke?) as
a brilliant light that was seen in the eastern sky for over 40 days.
\end{quote}

\begin{figure}[t]
\includegraphics[width=\linewidth]{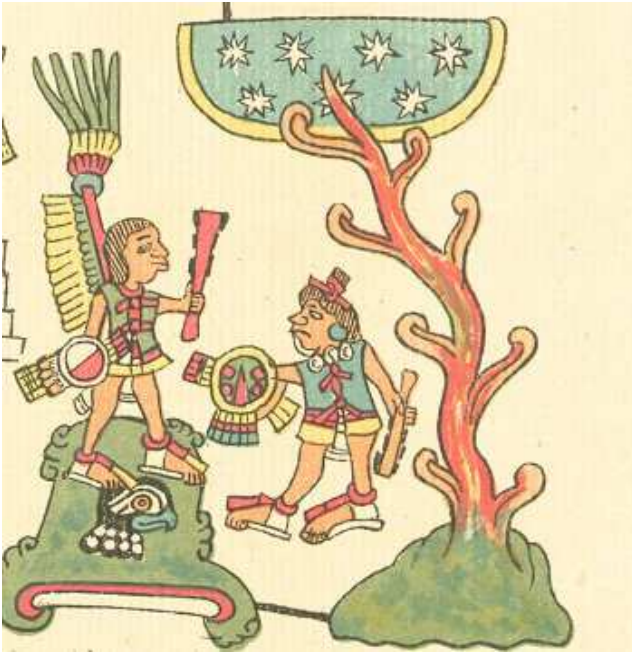}
\caption{Sketch from the \textit{Telleriano} tied to the year
    plaque of 1509.}
\label{fig:popocatepetl}
\end{figure}

The Figure \ref{fig:popocatepetl} clearly shows a column of red
flames rising from a mountain up to the starry sky.
A later annotation adds that the cloud banner preceded the return
of Quetzalcoatl whom the Aztecs seem to have associated with the
forthcoming arrival of Cortes.
Also, the chronology of Gallatin says for the corresponding year
that ``a great light was seen in the night towards the east'' and
it extended ``from the earth to the sky'' \cite{gallatin_1845}.
So, the sole allusion is some kind of a cloud or a light, but the
premonition of foreign people advancing must be a later
supplement after the knowledge of the existence of a man named
Cortes.
The phenomenon points to ash emission from an active volcano in
remote distance.
The mountain Popocatepetl is located 50 km to the southeast of
Tenochtitlan.
In the Nahuatl language of the Mexica the name translates
``smoking mountain''.
The suspicion of two strong eruptive episodes was expressed by
Martin-Del Pozzo, who believes that they took place in 1509 and
1519 each of which lasting for several months
\cite{martindelpozzo}.
Later Spanish writers merged the two incidents to create a sign
of possible disaster.
The inclusion of omens presaging an event reflects a re-working
of historical facts under late-medieval European influence.

In fact, the Aztec belief shows much weaker connections between
earthly processes (droughts, floods) and celestial ones (eclipses,
planetary positions).
Such is rather a custom of the astrological lore of Babylonians,
Chinese, Indians, and later Europeans.
In Mesoamerica there existed religious rituals, but it is difficult
to judge from the pictograms whether they were used as prayers, or
horoscopes, or prophecies for future events.
It is likely that the legendary style was imported as a post-conquest
feature to make ``sense'' of the historical present.
In this spirit, many rumours were created by the Conquistadores to
be part of a rationalisation of the Aztecs' defeat in order to
present Moctezuma as indecisive, powerless, and superstitious who
ultimately caused the fall of his own empire.
The early Spanish writers often portrayed their arrival as an
inevitable moment of history ordained by God.
The Europeans had substantial motives to log a ``seamless
transition'' of power, especially, for the authorities back home.


\section{Discussion}

The primary concern of this paper is a new list on solar eclipses
during the era of the Aztecs.
We replenished the stock of previous lists with additional sources,
debugged some errors, and present (hopefully) sound data including
a discussion to particular items.
Forty records from the various codices can be assigned to 23
eclipses.
32 entries (80\%) belong to the last 50 years before the conquest.
Only 19 accounts seem to agree with the correct year, the others
are shifted by one year give or take, and perhaps one or two are
doubtful.
The error of different years for the same eclipse could have arisen
from copying errors.
In general, the Aztec chroniclers present themselves as rather
careful timekeepers.
Our list of references for eclipses is far from complete and may
inspire others to search for more entries of this kind.
Along the way, we critically examined the astronomical knowledge
of the Aztecs.

We figured out that modern studies contain a lot of
overinterpretation.
The portraitures on the Aztec pictograms are widely debated, and
so are their derivative interpretations.
For example, suggestions for a given figure range from a god to a
demon to a ruler to the personification of celestial objects.
Modern authors draw an analogy between any of these options, and
the work ends up in different fictitious stories.
The fundament is spongy, but many go on constructing a theory on
``maybe''s and other speculations.
It seems hard for historians to admit that they just don't know
what the Aztec images mean.

The interest of the Aztecs in astronomy emerges quite low.
Though the calender is adopted from much earlier times
(Maya and other cultures), 
there is no sophisticated care for the sky.
Neither simple star maps are known, nor an accurate observation
of the moon's path.
There is no support to the view that the Aztecs made computations,
nor they discovered any cycles, nor tried to predict celestial
events like eclipses.
We find no evidence that they ever understood the scientific
basics for the cause of eclipses.
For example, there is no single mention that solar eclipses are
correlated with the moon.
The Aztecs just recorded about two dozen events with the recent
ones more frequent.
Nevertheless, eclipses were much feared but in a completely
different manner than prevalent in European or Mesopotamian
thought.
Their superstition did not concern the fate of a ruler or their
nation but the existence of the world as an entity.
According to their mythology the sun would suffer an earthquake
and perish in eternal darkness.

More insights into their belief are to be discovered, but we
recommend future studies to be based on safer grounds.
Astronomical events are precisely datable and can assist as an
independent method for verification of any theory.
The most severe mistakes, however, are made upon interpretation of
the textual evidence.


\section*{Acknowledgements}

This paper has its roots in Chapter 17.6 of the Habilitation
submitted to the University of Heidelberg, Germany, in February
2012.
The issue is thoroughly revised, expanded to additional items
and published ``as is'' without peer review.
The progress of this work was marked by tremendous obstacles.




\end{document}